\documentclass[preprint,aps,showpacs,nofootinbib,preprintnumbers,amsmath,amssymb]{revtex4-1}
\usepackage{epsfig}
\usepackage{subfigure}
\usepackage{dcolumn}% Align table columns on decimal point
\usepackage{bm}% bold math
\usepackage[usenames ,dvipsnames]{xcolor}
\usepackage{slashed}
\usepackage{graphicx,color}

\begin{document}
\title{
Four-body baryonic decays of  
$B\to p \bar{p} \pi^+\pi^-(\pi^+K^-)$ and 
$\Lambda \bar{p} \pi^+\pi^-(K^+K^-)$
}

\author{Y.K. Hsiao$^{1,2}$ and C.Q. Geng$^{1,2,3}$}
\affiliation{
$^{1}$Chongqing University of Posts \& Telecommunications, Chongqing, 400065, China\\
$^{2}$Department of Physics, National Tsing Hua University, Hsinchu, Taiwan 300\\
$^{3}$Synergetic Innovation Center for Quantum Effects and Applications (SICQEA),\\
 Hunan Normal University, Changsha 410081, China
}
\date{\today}

\begin{abstract}
We study the four-body baryonic $B\to {\bf B_1 \bar B_2}M_1 M_2$ decays with $\bf B_{1,2}$ ($M_{1,2}$) being charmless baryons (mesons). In accordance with the recent LHCb observations, each decay is considered to proceed through the $B\to M_1 M_2$ transition together with the production of a baryon pair. We obtain that ${\cal B}(B^-\to \Lambda\bar p \pi^+\pi^-)=(3.7^{+1.5}_{-1.0} )\times 10^{-6}$ and ${\cal B}(\bar B^0\to p\bar p \pi^+\pi^-,p\bar p \pi^+ K^-)=(3.0\pm 0.9,6.6\pm 2.4)\times 10^{-6}$, in agreement with the data. We also predict ${\cal B}(B^-\to\Lambda\bar p K^+ K^-)=(3.0^{+1.3}_{-0.9})\times 10^{-6}$, which is accessible to the LHCb and BELLE experiments.
\end{abstract}
%\pacs{}

\maketitle
\section{introduction}
One of the main purposes of the $B$ factories and current LHCb is to study CP violation (CPV), 
which is important for us to understand the puzzle of the matter-antimatter asymmetry
in the Universe. As the observables, the (in)direct CP-violating asymmetries (CPAs) require 
both weak and strong phases~\cite{Geng:2016kjv,Hsiao:2014mua,Geng:2006jt},
whereas the T-violating triple momentum product correlations (TPCs), such as
$\vec{p}_1\cdot (\vec{p}_2\times \vec{p}_3)$ in a four-body decay,
do not necessarily need a strong phase~\cite{Bensalem,Geng:2005wt}.
For example, 
the LHCb Collaboration has provided the first evidence for CPV from the TPCs 
in $\Lambda_b\to p\pi^-\pi^+\pi^-$~\cite{Aaij:2016cla}, and measured TPCs
in $\Lambda_b\to p K^-\mu^+\mu^-$~\cite{Aaij:2017mib}.
As the similar baryonic cases, 
the four-body baryonic $B$ decays can also provide TPCs.

\begin{figure}[t]
\centering
\includegraphics[width=2.1in]{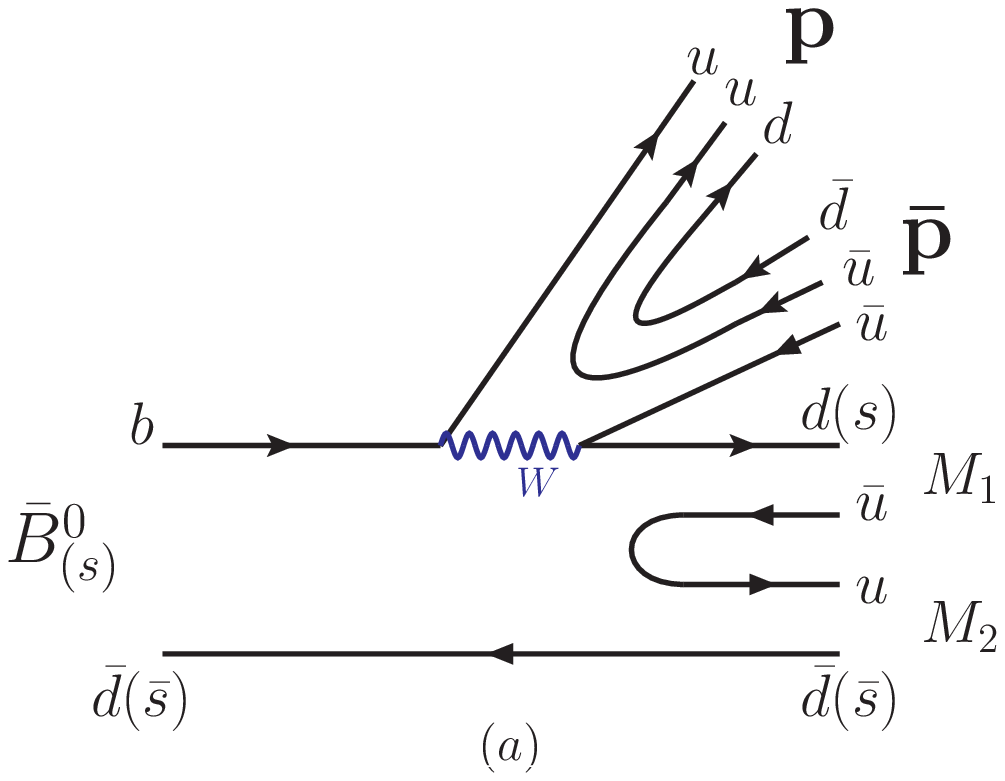}
\includegraphics[width=2.1in]{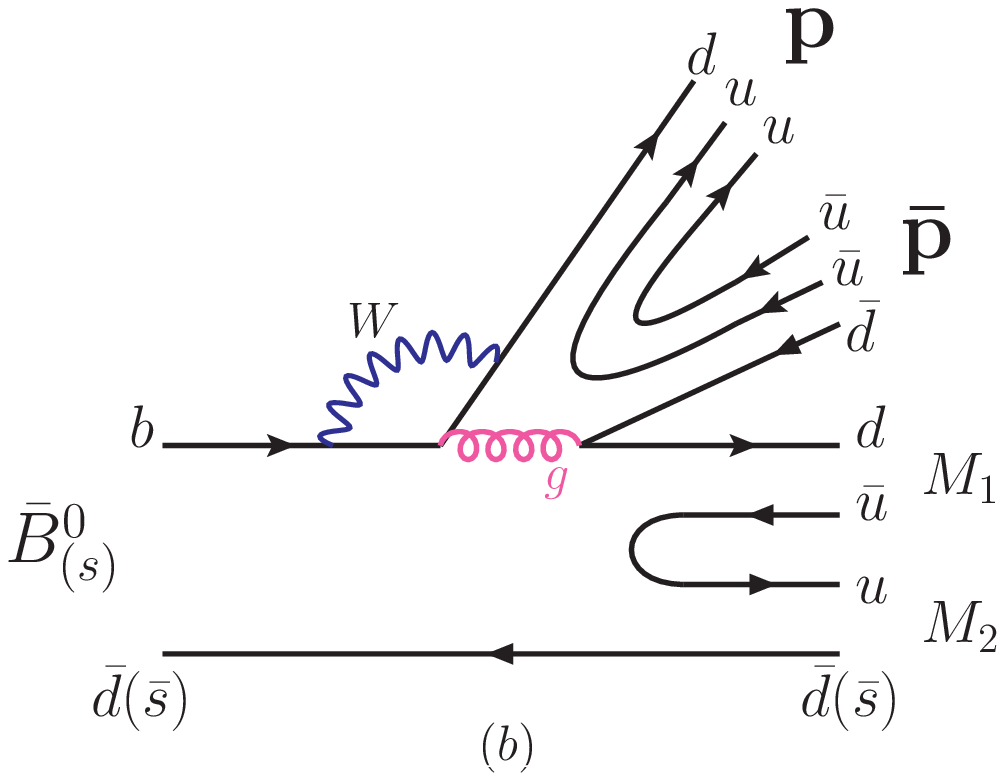}
\includegraphics[width=2.1in]{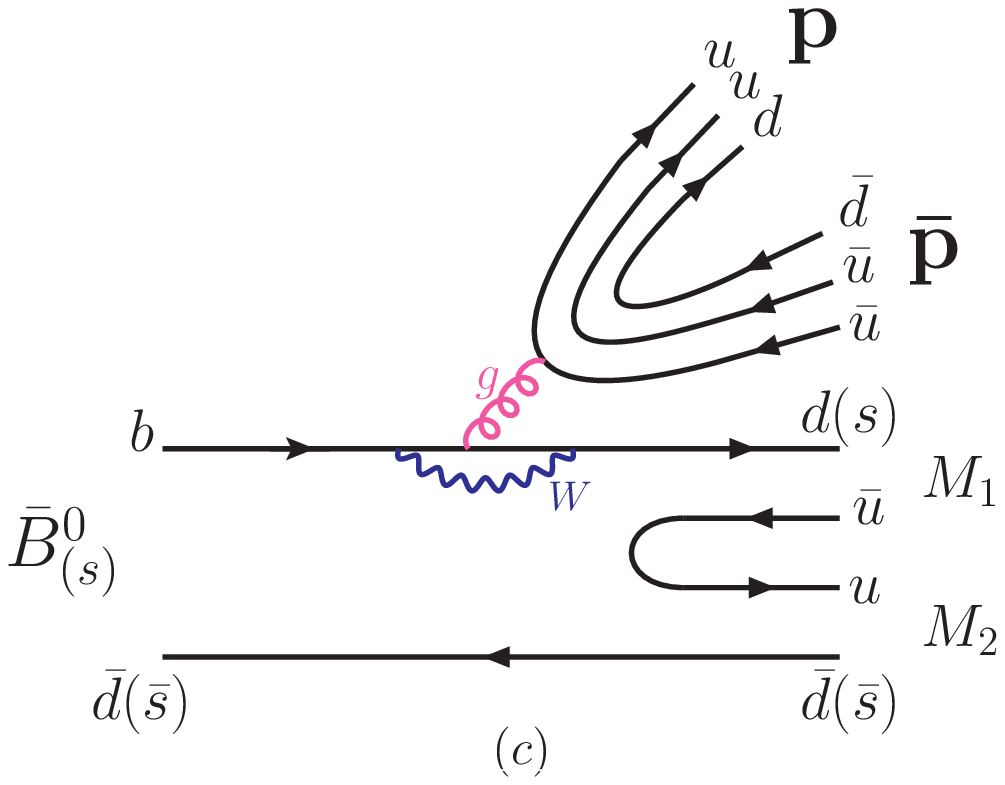}
\includegraphics[width=2.1in]{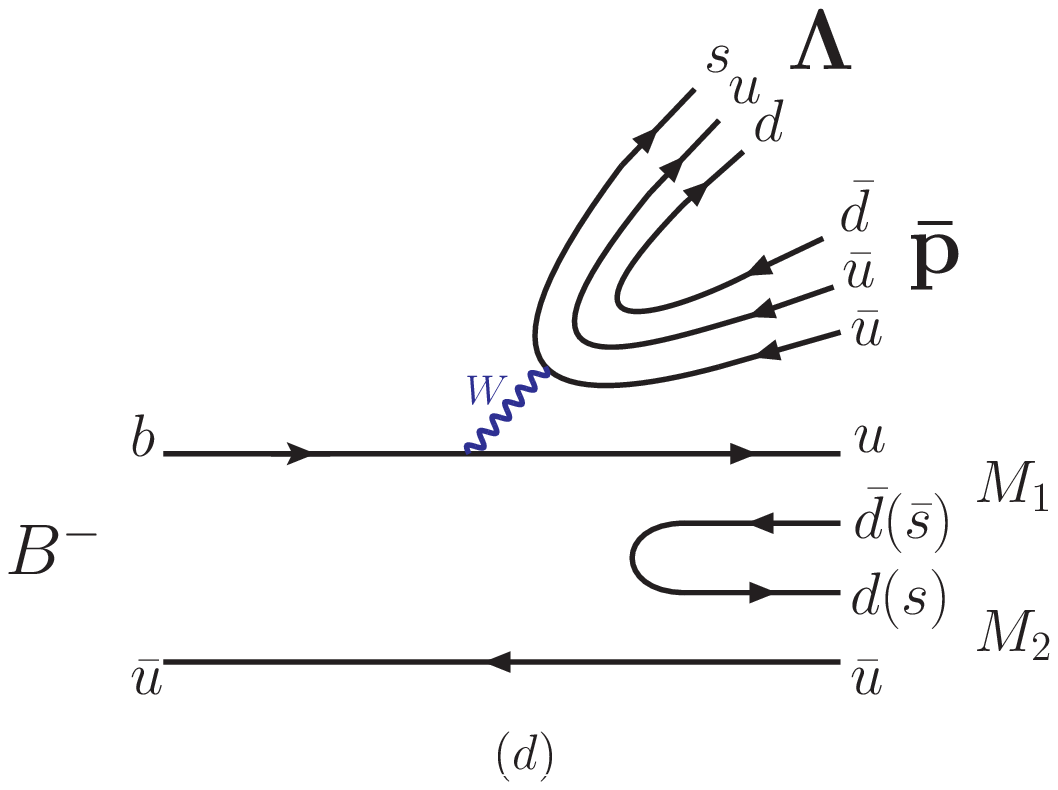}
\includegraphics[width=2.1in]{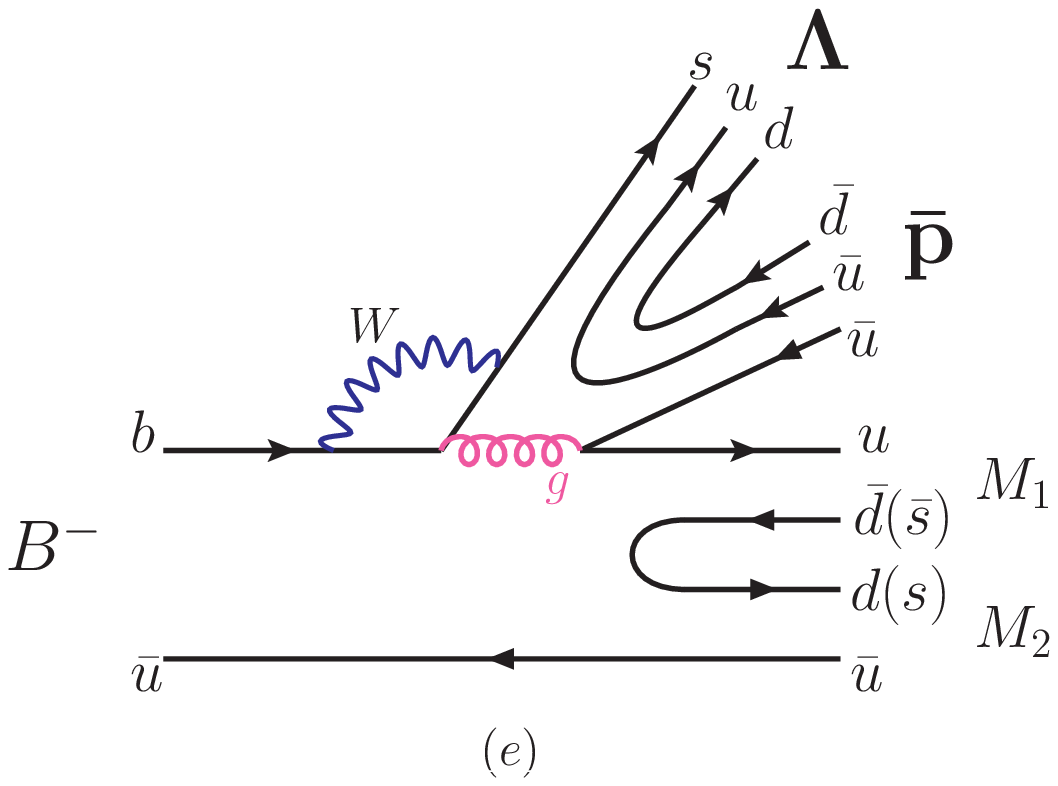}
\caption{Feynman diagrams for the charmless four-body baryonic $B$ decays, where
(a,b,c) depict the $\bar B^0_{(s)}\to p\bar p M_1M_2$ decays,
while (d,e) the $B^-\to \Lambda\bar p M_1M_2$ decays.}\label{dia}
\end{figure}

For a long time, the $B^-\to \Lambda \bar p \pi^+\pi^-$ decay
was the  only observed decay mode 
in $B\to{\bf B_1\bar B_2}M_1M_2$~\cite{Chen:2009xg}. 
Until very recently,  more four-body baryonic  B decays have been 
%measured by the LHCb~\cite{BtoBBMM_LHCb},
observed by the LHCb~\cite{BtoBBMM_LHCb},
which motivate us to give  theoretical estimations on the corresponding decay branching ratios.
The experimental measurements for the branching ratios of 
$\bar B^0/B^-\to{\bf B_1\bar B_2}M_1M_2$ 
at the level of $10^{-6}$ are given by~\cite{Chen:2009xg,BtoBBMM_LHCb}
\begin{eqnarray}\label{br_data}
{\cal B}(\bar B^0\to p \bar p \pi^+\pi^-)
&=&(3.0\pm 0.2\pm 0.2\pm 0.1)\times 10^{-6}\,,\nonumber\\
{\cal B}(\bar B^0\to p \bar p K^\mp\pi^\pm)
&=&(6.6\pm 0.3\pm 0.3\pm 0.3)\times 10^{-6}\,,\nonumber\\
{\cal B}(B^-\to \Lambda \bar p \pi^+\pi^-)
&=&(5.92^{+0.88}_{-0.84}\pm 0.69)\times 10^{-6}\,,
\end{eqnarray}
where the resonant ${\cal B}(B^-\to \Lambda \bar p$ 
$(\rho^0,f_2(1270)\to)\pi^+\pi^-)$
have been excluded from the data~\cite{Chen:2009xg}.
In comparison with
${\cal B}(\bar B^0\to p \bar p K^+ K^-)
\simeq (1.3\pm 0.3)\times 10^{-7}$ and 
${\cal B}(\bar B^0_s\to p\bar p \pi^+ \pi^-)
%\simeq (4.6\pm 2.0)\times 10^{-7}$~\cite{BtoBBMM_LHCb},
<7.3\times 10^{-7}$ (90\% C.L.)~\cite{BtoBBMM_LHCb},
the decays with ${\cal B}\sim10^{-6}$ in Eq.~(\ref{br_data}) 
are recognized to have the same theoretical correspondence,
where $\bar B^0/B^-\to{\bf B_1\bar B_2}M_1M_2$
proceed through the $B\to M_1M_2$ transition along with 
the $\bf B_1\bar B_2$ production, as depicted in Fig.~\ref{dia}.
Note that
the $\bar B^0_s$ decays of $\bar B^0_s\to p\bar p K^\pm\pi^\mp$ and $p\bar p K^+ K^-$
with $\bar s$ being replaced by $\bar d$ 
in $\bar B^0\to p \bar p \pi^+\pi^-$ and $p \bar p K^\mp\pi^\pm$
 have also been found with the branching ratios of order $10^{-6}$~\cite{BtoBBMM_LHCb}, respectively.

In this report, 
we will calculate the four-body baryonic $B$ decays in accordance with
the decaying processes in Fig.~\ref{dia},
with the extraction of 
the $B\to M_1 M_2$ transition form factors from the $B\to D^{(*)}M_1M_2$ and 
$B\to M_1M_2M_3$ decays and the adoption of the timelike baryonic form factors 
from the two-body and three-body baryonic $B$ decays.
Our theoretical approach will be useful for the estimations of 
TPCs in $B\to{\bf B_1\bar B_2}M_1M_2$ to be compared to
%the future measurements at the LHCb.
 future measurements by the LHCb.

%=======================
\vspace*{0.3cm}
\vspace*{0.3cm}
%======================
\section{Formalism}
In terms of the quark-level effective Hamilontion for the charmless $b\to q_1\bar q_2 q_3$ transition, 
the amplitudes of the four-body baryonic $B$ decays 
by the generalized factorization approach are derived as~\cite{ali} 
\begin{eqnarray}\label{amp1}
&&{\cal A}_1(\bar B^0_{(s)}\to p\bar p M_1M_2)=
\frac{G_F}{\sqrt 2}\bigg\{\bigg[
\langle p\bar p|\alpha_+^q(\bar u u)_V-\alpha_-^q(\bar u u)_A|0\rangle+
\langle p\bar p|\beta_+^q(\bar dd)_V-\beta_-^q(\bar dd)_A|0\rangle\nonumber\\
&&+(\alpha_4^q-\alpha_{10}^q/2)\langle p\bar p|(\bar qq)_{V-A}|0\rangle
\bigg]\langle M_1M_2|(\bar q b)_{V-A}|\bar B^0_{(s)}\rangle\nonumber\\
&&+\alpha_6^q\langle p\bar p|(\bar qq)_{S+P}|0\rangle
\langle M_1M_2|(\bar qb)_{S-P}|\bar B^0_{(s)}\rangle\bigg\}\,,
\nonumber\\
&&{\cal A}_2(B^-\to\Lambda\bar p M_1 M_2)=%\nonumber\\&&
\frac{G_F}{\sqrt 2}\bigg\{
(\alpha_1^s+\alpha_4^s)\langle \Lambda\bar p|(\bar s u)_{V-A}|0\rangle 
\langle M_1M_2|(\bar u b)_{V-A}|B^-\rangle\nonumber\\
&&+\alpha_6^s\langle \Lambda\bar p|(\bar s u)_{S+P}|0\rangle
\langle M_1 M_2|(\bar u b)_{S-P}|B^-\rangle\bigg\}\,,
\end{eqnarray}
where  $G_F$ is the Fermi constant, $V_{ij}$ are the CKM matrix elements,
and $(\bar q_1 q_2)_{V(A)}$ and $(\bar q_1 q_2)_{S(P)}$ stand for 
$\bar q_1 \gamma_\mu(\gamma_5) q_2$ and $\bar q_1(\gamma_5) q_2$, respectively.
The parameters $\alpha^q_{\xi}$ and $\beta^q_\eta$ 
in Eq.~(\ref{amp1})
are given by
\begin{eqnarray}
\alpha^q_\pm&=&\alpha_2^q+\alpha_3^q\pm\alpha_5^q+\alpha_9^q\,,
\beta^q_\pm=\alpha_3^q\pm\alpha_5^q-\alpha_9^q/2\,,\nonumber\\
\alpha_{1,2}^q&=&V_{ub}V_{uq}^* a_{1,2} \,,
\alpha_j^q=-V_{tb}V_{tq}^*a_j\,,
\alpha_6^q=V_{tb}V^*_{tq}2a_6\,,
\end{eqnarray}
with $q=(d,s)$ and $j=(3,4,5,9,10)$,
where $a_i\equiv c^{eff}_i+c^{eff}_{i\pm 1}/N_c^{eff}$ for $i=$ odd (even)
with the effective color number $N_c^{eff}$ and
Wilson coefficients $c_{i}^{eff}$ in Ref.~\cite{ali}.
From 
${\cal A}_1(\bar B^0_{(s)}\to p\bar p M_1M_2)$ and 
${\cal A}_2(B^-\to\Lambda\bar p M_1 M_2)$ in Eq.~(\ref{amp1}), 
the allowed decays are
\begin{eqnarray}
&&
\bar B^0\to p\bar p \pi^+\pi^-,\,
\bar B^0_s\to p\bar p K^+\pi^-\,,\;\text{(q=d)}\nonumber\\ 
&&
\bar B^0\to p\bar p \pi^+ K^-,\,
\bar B^0_s\to p\bar p K^+ K^-\,,\;\text{(q=s)}\nonumber\\ 
&&
B^-\to \Lambda\bar p \pi^+\pi^-,\;B^-\to\Lambda\bar p K^+ K^-\,.
\end{eqnarray}
Note that  
the $\bar B^0\to p\bar p \pi^+ K^-$ and $\bar B^0_s\to p\bar p K^+ K^-$ decays
have the matrix elements of
$\langle p\bar p|(\bar s s)_{V,A,S,P}|0\rangle$ with the $\bar ss$ quark currents,
which eventually cause the terms of $\alpha_{4,6,10}^s$
to give nearly zero contributions
due to the OZI suppression of $\bar s s\to p\bar p$~\cite{Hsiao:2014tda}.

For the matrix elements in Eq.~(\ref{amp1}), the baryon-pair productions
from the quark currents are given by~\cite{Chua:2002yd,Geng:2005wt} 
\begin{eqnarray}\label{FFactor1}
\langle {\bf B_1\bar B_2}|\bar q_1\gamma_\mu q_2|0\rangle
&=&
\bar u\bigg[F_1\gamma_\mu+\frac{F_2}{m_{\bf B_1}+m_{\bf \bar B_2}}i\sigma_{\mu\nu}q_\mu\bigg]v\;,\nonumber\\
\langle {\bf B_1\bar B_2}|\bar q_1\gamma_\mu \gamma_5 q_2|0\rangle
&=&\bar u\bigg[g_A\gamma_\mu+\frac{h_A}{m_{\bf B_1}+m_{\bf \bar B_2}}q_\mu\bigg]\gamma_5 v\,,\nonumber\\
\langle {\bf B_1\bar B_2}|\bar q_1 q_2|0\rangle &=&f_S\bar uv\;,
\langle {\bf B_1\bar B_2}|q_1\gamma_5 q_2|0\rangle =g_P\bar u \gamma_5 v\,,
\end{eqnarray}
where $q=p_{\bf B_1}+p_{\bf\bar B_2}$, $t\equiv q^2$,
$u$($v$) is the (anti-)baryon spinor, and  
$(F_{1,2},g_A,h_A,f_S,g_P)$ are the timelike baryonic form factors.
On the other hand, the $B\to M_1 M_2$ transition matrix elements 
are parameterized as~\cite{Lee:1992ih}
\begin{eqnarray}\label{BtoMM}
&&\langle M_1 M_2|\bar q_1\gamma_\mu(1-\gamma_5)b|B\rangle\nonumber\\
&=&
h\epsilon_{\mu\nu\alpha\beta}p_B^\nu p^\alpha (p_{M_2}-p_{M_1})^\beta
%\langle M_1 M_2|\bar q_1\gamma_\mu\gamma_5 b|B\rangle&=&
%-irq_\mu-iw_+ p_\mu-iw_-(p_{M_2}-p_{M_1})\,,
+irq_\mu+iw_+ p_\mu+iw_-(p_{M_2}-p_{M_1})\,,
\end{eqnarray}
where $p=p_{M_2}+p_{M_1}$ and $(h,r,w_{\pm})$ are the form factors. 
Subsequently, one can also get 
$\langle M_1 M_2|\bar q_1(\gamma_5)b|B\rangle$ from Eq.~(\ref{BtoMM}) 
based on equations of motion. In terms of the approach of pQCD counting rules,
the momentum dependences for
the $0\to {\bf B_1\bar B_2}$ and $B\to M_1M_2$
transition form factors are 
given by~\cite{Brodsky:1973kr,Brodsky:2003gs,Chua:2002pi,Chua:2004mi}
\begin{eqnarray}\label{timelikeF2}
&&F_1=\frac{\bar C_{F_1}}{t^2}\,,\;g_A=\frac{\bar C_{g_A}}{t^2}\,,\;
f_S=\frac{\bar C_{f_S}}{t^2}\,,\;g_P=\frac{\bar C_{g_P}}{t^2}\,,\;\nonumber\\
&&h=\frac{C_h}{t^2}\,,\;w_-=\frac{D_{w_-}}{t^2}\,,
\end{eqnarray}
where $\bar C_i=C_i [\text{ln}({t}/{\Lambda_0^2})]^{-\gamma}$
with $\gamma=2.148$ and $\Lambda_0=0.3$ GeV.
We note that since
$F_2$ is derived to be $F_2=F_1/(t\text{ln}[t/\Lambda_0^2])$~\cite{Belitsky:2002kj},
which is much less than $F_1$, while 
the small value of ${\cal B}(\bar B^0\to p\bar p)=
(1.5^{+0.7}_{-0.5})\times 10^{-8}$~\cite{pdg,Aaij:2013fta}
causes a tiny $C_{h_A}$~\cite{Hsiao:2014zza} in $h_A=C_{h_A}/t^2$,
we may not consider the effects from $F_2$ and $h_A$.
In addition, by following Ref.~\cite{Chua:2002pi},
we have neglected the terms related to
$r$ and 
$w_+$ in Eq.~(\ref{BtoMM}) 
due to the wrong parity~\cite{Drutskoy:2002ib}.
%fact that
%$r$ and $w_+$ present to be symmetric
%by the reordering $p_{M_1}$ and $p_{M_2}$, indicating even parity,
%whereas the angular analysis of the $B\to D^{(*)}KK$ decays~\cite{Drutskoy:2002ib}
%favor $J^P=1^-$ as the quantum numbers of the meson pair. 

%========================
\begin{figure}[t]
\centering
\includegraphics[width=2.8in]{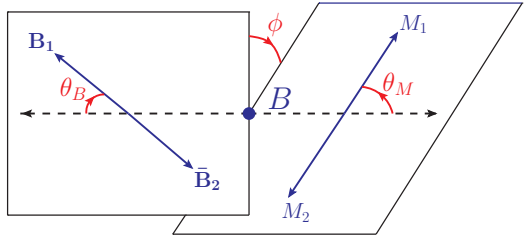}
\caption{Three angles of $\theta_{\bf B}$, $\theta_{\bf M}$,
and $\phi$ in the phase space for the four-body $B\to {\bf B_1\bar B_2}M_1M_2$ decays.
}\label{4body}
\end{figure}
%========================
The integration over the phase space of the four-body 
$B(p_B)\to {\bf B_1}(p_{\bf B_1}){\bf\bar B_2}(p_{\bf \bar B_2})$
$M_1(p_{M_1}) M_2 (p_{M_2})$ decay relies on
the five kinematic variables, that is, $s\equiv p^2$, 
$t$ and the three angles of $\theta_{\bf B}$, $\theta_{\bf M}$ and $\phi$.
In Fig.~\ref{4body},
the angle $\theta_{\bf B(M)}$ is between $\vec{p}_{\bf B_1}$  ($\vec{p}_{M_1}$)
of the $\bf B_1\bar B_2$ ($M_1 M_2$) rest frame and 
the line of flight of the $\bf B_1\bar B_2$ ($M_1 M_2$) system in the $B$ meson rest frame, 
while the angle $\phi$ is from the $\bf B_1\bar B_2$ plane 
to the $M_1 M_2$ plane,
defined by the momenta of the $\bf B_1\bar B_2$ and $M_1 M_2$ pairs
in the $B$ rest frame, respectively.
The partial decay width reads~\cite{Geng:2011tr,Geng:2012qn}
\begin{eqnarray}
d\Gamma=\frac{|\bar {\cal A}|^2}{4(4\pi)^6 m_B^3}X
\alpha_{\bf B}\alpha_{\bf M}\, ds\, dt\, d\text{cos}\,\theta_{\bf B}\, d\text{cos}\,\theta_{\bf M}\, d\phi\,,
\end{eqnarray}
where $X$, $\alpha_{\bf B}$ and $\alpha_{\bf M}$ are given by
\begin{eqnarray}
X&=&\bigg[\frac{1}{4}(m_B^2-s-t)^2-st\bigg]^{1/2}\,,\nonumber\\
\alpha_{\bf B}&=&\frac{1}{t}\lambda^{1/2}(t,m_{\bf B_1}^2,m_{\bf \bar B_2}^2)\,,\nonumber\\
\alpha_{\bf M}&=&\frac{1}{s}\lambda^{1/2}(s,m_{M_1}^2,m_{M_2}^2)\,,
\end{eqnarray}
respectively,
with $\lambda(a,b,c)=a^2+b^2+c^2-2ab-2bc-2ca$, while
the allowed ranges of the five variables are given by
\begin{eqnarray}
&&(m_{M_1}+m_{M_2})^2\leq s\leq (m_{B}-\sqrt{t})^2\,,\;\;
(m_{\bf B_1}+m_{\bf \bar B_2})^2\leq t\leq (m_B-m_{M_1}-m_{M_2})^2\,,\nonumber\\
&&0\leq \theta_{\bf B},\,\theta_{\bf M}\leq \pi\,,\;\;0\leq \phi\leq 2\pi\,.
\end{eqnarray}

\section{Numerical Results and Discussions }
For the numerical analysis, 
the CKM matrix elements in the Wolfenstein parameterization 
are presented as
\begin{eqnarray}
&&(V_{ub},\,V_{tb})=(A\lambda^3(\rho-i\eta),1)\,,\nonumber\\
&&(V_{ud},\,V_{td})=(1-\lambda^2/2,A\lambda^3)\,,\nonumber\\
&&(V_{us},\,V_{ts})=(\lambda,-A\lambda^2),
\end{eqnarray}
with $(\lambda,\,A,\,\rho,\,\eta)=(0.225,\,0.814,\,0.120\pm 0.022,\,0.362\pm 0.013)$~\cite{pdg}.
To estimate the non-factorizable effects
in the generalized factorization approach~\cite{ali},
$N_c^{eff}$ ranges from 2 to $\infty$. In Table~\ref{alpha_i},
we show the values of $a_i$
for the $b\to d$ and $b\to s$ transitions with $N_c^{eff}=(2, 3,\infty)$, respectively.
%=================
\begin{table}[b]
\caption{The parameters $a_i$ with $N_c^{eff}=2,\,3$, and $\infty$
to estimate the non-factorizable effects in the generalized factorization.}\label{alpha_i}
{\footnotesize
\begin{tabular}{|c||ccc|ccc|}
\hline
&\multicolumn{3}{c|}{($b\to d$ transition)}&\multicolumn{3}{c|}{($b\to s$ transition)}\\
$a_i$& $N_c^{eff}=2$ &  $3$ & $\infty$ &$N_c^{eff}=2$&$3$&$\infty$\\%\hline
\cline{2-7}
$a_1$ 
& ----- &  ----- & -----
&$0.98$&$1.05$&$1.17$ \\
$a_2$ 
&$0.22$&$0.02$&$-0.37$
&$0.22$&$0.02$&$-0.37$ \\
$10^{4}a_3$ 
&$-10.4- 6.9i$  &  $72.4$ & $237.9+ 13.9i$
&$-13.1-15.6i$ &  $72.4$ & $243.2+31.2i$
\\
$10^{4}a_4$ 
& $-377.6-34.7i$ &  $-417.2-37.0i$ & $-496.5-41.6i$ 
& $-391.0-77.9i$ &  $-431.6-83.1i$ & $-512.6-93.5i$
\\
$10^{4}a_5$ 
& $-171.4 - 6.9i$ &  $-65.8$ & $145.3+13.9i$
& $-174.1-15.6i$ &  $-65.8$ & $150.7+31.2i$
\\
$10^{4}a_6$ 
& $-560.7-34.7i$ &  $-584.9-37.0i$ & $-633.4-41.6i$
& $-574.1-77.9i$ &  $-599.3-83.1i$ & $-649.5-93.5i$
\\
$10^{4}a_9$ 
& $-93.3-1.4i$ & $-99.5-1.4i$ & $-112.0-1.4i$
& $-93.5-2.2i$ & $-99.8-2.2i$ & $-112.3-2.2i$
\\
$10^{4}a_{10}$ 
&$-18.5-0.7i$&$0.18-0.46i$&$37.5$
& ----- &  ----- & -----\\
\hline
\end{tabular}
}
\end{table}
%=================

\begin{figure}[t]
\centering
\includegraphics[width=2.3in]{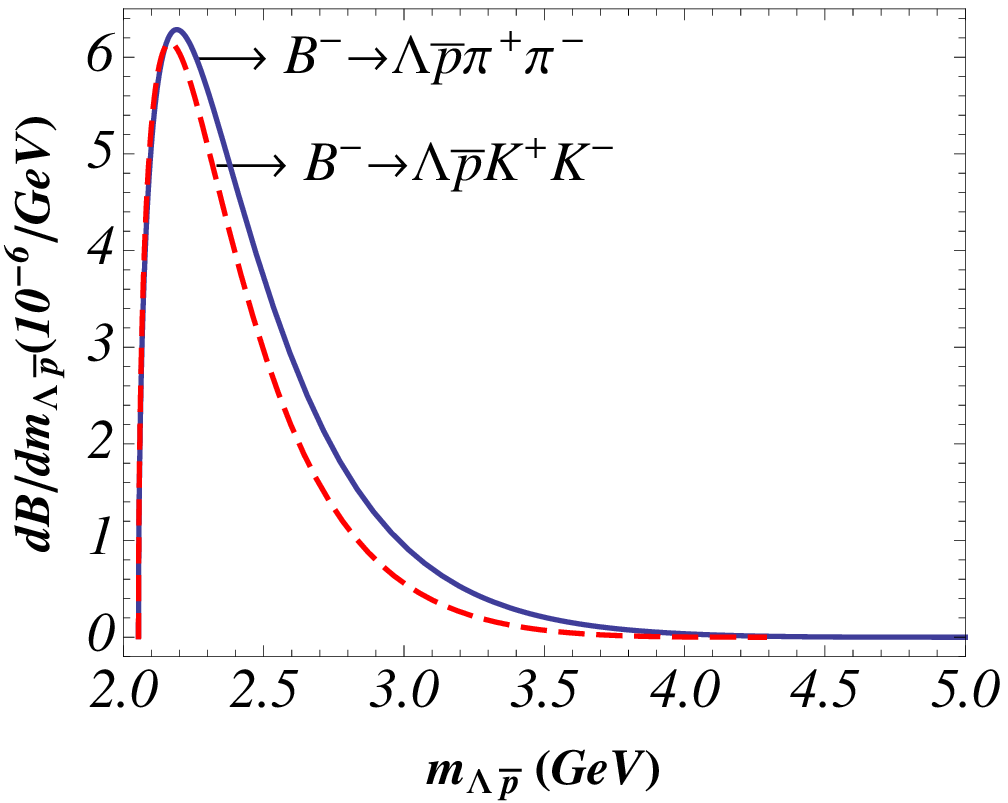}
\includegraphics[width=2.3in]{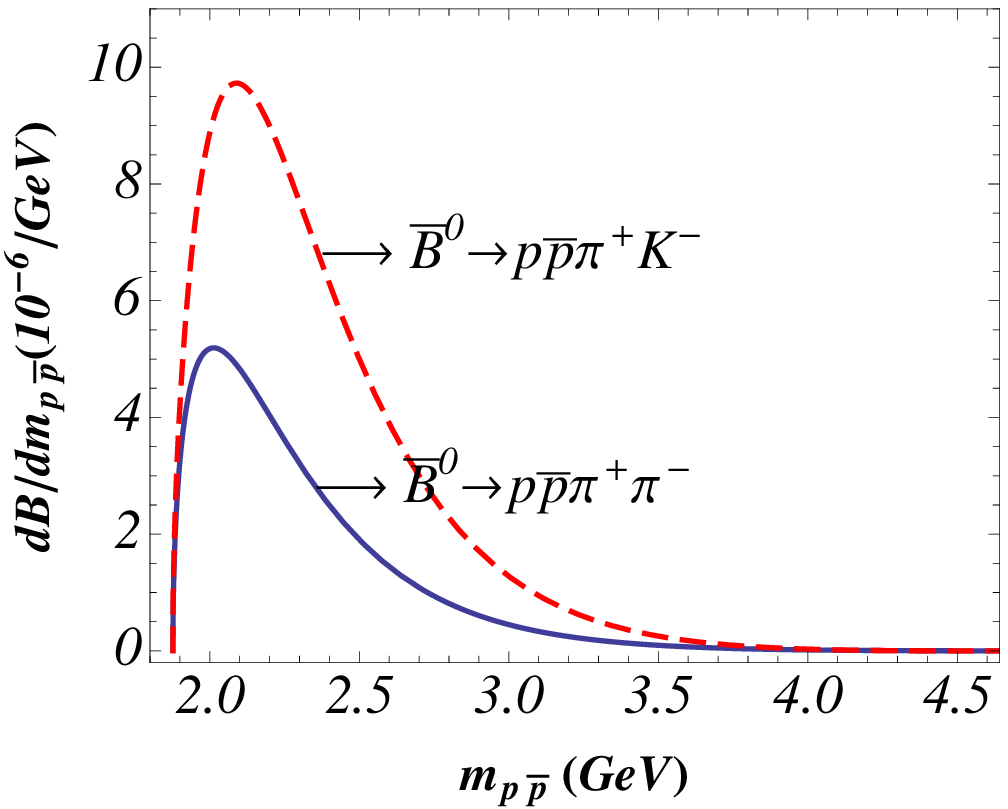}
\caption{Invariant dibaryon mass spectra for 
$B^-\to \Lambda\bar p M_1M_2$ (left panel) and 
$\bar B^0\to p\bar p M_1M_2$ (right panel), respectively.}\label{spec}
\end{figure}

According to the extractions of $(C_h,C_{w_-})$ in Refs.~\cite{Chua:2002pi,Chua:2004mi},
we fit the $B\to \pi\pi$ transition form factors with the branching ratios of 
$\bar B^0\to D^{(*)0} \pi^+ \pi^-$, 
$B^-\to \pi^-\pi^+\pi^-$ and
$B^-\to K^{*-}\pi^+\pi^-$,
and the $B\to (K\pi,KK)$ ones with those of 
$B^-\to D^{(*)0} K^- K^0$, 
$\bar B^0\to D^0 K^- \pi^+$ and
$B^-\to K^{*-}K^+ K^-$. Note that 
the contributions from the resonant
$B\to D^{(*)}(M_0\to)M_1M_2$, and
$B^-\to K^{*-} (M_0\to)M_1 M_2$ decays
with $\rho^0,f_2(1270)\to \pi^+\pi^-$ or $\phi\to K^+K^-$
have been excluded from the data. 
Unfortunately, the current observations of 
${\cal B}(\bar B^0_s\to M_1 M_2 M_3)$ are not sufficient
for us to extract the $\bar B^0_s\to M_1 M_2$ transition form factors.
As a result, we obtain
\begin{eqnarray}
(C_h,C_{w_-})|_{B\to \pi\pi}&=&(3.6\pm 0.3,0.7\pm 0.2)\,GeV^3\,, \nonumber\\
(C_h,C_{w_-})|_{B\to KK(K\pi)}&=&(-38.9\pm 3.3,14.2\pm 2.3)\,GeV^3\,.
\end{eqnarray}
The timelike baryonic form factors in Eq.~(\ref{FFactor1})
can be related with the $SU(3)$ flavor and $SU(2)$ spin symmetries,
such that $(C_{F_1},C_{g_A},C_{f_S},C_{g_P})$ are recombined 
by a new set of constant parameters
as~\cite{Brodsky:1973kr,Chua:2002yd,Geng:2016fdw,Hsiao:2016amt,ang_BtopLpi}
\begin{eqnarray}\label{0toBB}
&&
C_{F_1}=\frac{5}{3}C_{||}+\frac{1}{3}C_{\overline{||}}\,,\;
C_{g_A}=\frac{5}{3}C_{||}^*-\frac{1}{3}C_{\overline{||}}^*\,,\;\;
\text{(for $\langle p\bar p|\bar u\gamma_\mu(\gamma_5)u|0\rangle$)}\nonumber\\
&&
C_{F_1}=\frac{1}{3}C_{||}+\frac{2}{3}C_{\overline{||}}\,,\;
C_{g_A}=\frac{1}{3}C_{||}^*-\frac{2}{3}C_{\overline{||}}^*\,,\;\;
\text{(for $\langle p\bar p|\bar d\gamma_\mu(\gamma_5)d|0\rangle$)}\nonumber\\
&&
C_{f_S}=\frac{1}{3}\bar C_{||}\,,\;C_{g_P}=\frac{1}{3}\bar C_{||}^*\,,\;\;
\text{(for $\langle p\bar p|\bar d(\gamma_5)d|0\rangle$)}\nonumber\\
&&
C_{F_1}=\sqrt\frac{3}{2}C_{||}\,,\;
C_{g_A}=\sqrt\frac{3}{2}C_{||}^*\,,\;\;
\text{(for $\langle \Lambda\bar p|\bar s\gamma_\mu(\gamma_5)u|0\rangle$)}\nonumber\\
&&
C_{f_S}=-\sqrt\frac{3}{2}\bar C_{||}\,,
C_{g_P}=-\sqrt\frac{3}{2}\bar C_{||}^*\,,\;\;
\text{(for $\langle \Lambda\bar p|\bar s(\gamma_5)u|0\rangle$)}%\nonumber\\
\end{eqnarray}
with  
$C_{||(\overline{||})}^*\equiv C_{||(\overline{||})}+\delta C_{||(\overline{||})}$ and 
$\bar C_{||}^*\equiv \bar C_{||}+\delta \bar C_{||}$, in which 
$\delta C_{||(\overline{||})}$ and $\delta \bar C_{||}$ have been added 
to explain the large and unexpected 
angular distributions in $\bar B^0\to \Lambda\bar p\pi^+$ and 
$B^-\to \Lambda\bar p\pi^0$~\cite{ang_BtopLpi,Wang:2007as},
to account for the fact that the $SU(3)$ flavor and $SU(2)$ spin symmetries 
at large $t$ ($t\to \infty$)~~\cite{Brodsky:1973kr} should be broken 
at $t\simeq m_B^2$~\cite{ang_BtopLpi}. 
The extractions of the form factors by the data of 
$\bar B^0\to n\bar p D^{*+}$, 
$\bar B^0\to \Lambda\bar p D^{(*)+}$,
$\bar B^0\to \Lambda\bar p \pi^+$, 
$B^-\to \Lambda\bar p  (\pi^0,\rho^0)$, 
$\bar B^0_{(s)}\to p\bar p$ and
$B^-\to \Lambda\bar p $
give~\cite{Geng:2016fdw} 
\begin{eqnarray}\label{fitC2}
&&(C_{||},\,\delta C_{||})=(154.4\pm 12.1,\,19.3\pm 21.6)\;{\rm GeV}^{4}\,,\nonumber\\
&&(C_{\overline{||}},\,\delta C_{\overline{||}})=(18.1\pm 72.2,\,-477.4\pm 99.0)\;{\rm GeV}^{4}\,,\nonumber\\
&&(\bar C_{||},\,\delta\bar C_{||})=(537.6\pm 28.7,\,-342.3\pm 61.4)\;{\rm GeV}^{4}\,,
\end{eqnarray}
where the added constants for the broken effects have 
been approved by the excellent agreement for 
${\cal B}(\bar B^0_s\to \Lambda \bar p K^+ +\bar \Lambda p K^-)$~\cite{LHCb:2017khw}.
Subsequently, 
we evaluate the branching ratios of $B\to {\bf B_1\bar B_2}M_1 M_2$
as shown in Table~\ref{br_BtoBBMM}, and draw the distributions vs. $m_{\bf B_1\bar B_2}$
in Fig.~\ref{spec}.
%=======================
%\vspace*{0.3cm}
%\vspace*{0.3cm}
%======================

%=======================
\begin{table}[b]%[htb]
\caption{The branching ratios of $B\to{\bf B_1\bar B_2}M_1M_2$,
where the errors come from the non-factorizable effects, 
CKM matrix elements, and form factors, respectively.}\label{br_BtoBBMM}
\begin{tabular}{|c|c|c|}
\hline
branching ratios&our results&data\\\hline
$10^6{\cal B}(B^-\to \Lambda\bar p \pi^+\pi^-)$
&$3.7^{+1.2}_{-0.5}\pm 0.1\pm 0.9$&$5.9\pm 1.1$\\
$10^6{\cal B}(B^-\to\Lambda\bar p K^+ K^-)$
&$3.0^{+1.1}_{-0.5}\pm 0.1\pm 0.7$&-----\\
$10^6{\cal B}(\bar B^0\to p\bar p \pi^+\pi^-)$
&$3.0^{+0.5}_{-0.3}\pm 0.3\pm 0.7$&$3.0\pm 0.3$\\
$10^6{\cal B}(\bar B^0\to p\bar p \pi^\pm K^\mp)$
&$6.6\pm 0.5\pm 0.0\pm 2.3$&$6.6\pm 0.5$\\
\hline
\end{tabular}
\end{table}
%============================

As seen in Table~\ref{br_BtoBBMM}, although the predicted result of
${\cal B}(B^-\to \Lambda\bar p \pi^+\pi^-)=(3.7^{+1.5}_{-1.0} )\times 10^{-6}$
is a little lower, it is consistent with the data in Eq.~(\ref{br_data}) 
by taking the uncertainties into account. 
With the replacement of $B^-\to \pi^+\pi^-$ by $B^-\to K^+ K^-$,
the $B^-\to \Lambda\bar p \pi^+\pi^-$ and $\Lambda\bar p K^+K^-$ decays
share the same decaying configuration. We hence predict that 
${\cal B}(B^-\to\Lambda\bar p K^+ K^-)=(3.0^{+1.3}_{-0.9})\times 10^{-6}$,
which is accessible to the LHCb and BELLE experiments.
Unlike the $B^-\to \Lambda\bar p M_1M_2$ decays, where 
$a_{1,4,6}$ are stable by ranging $N_c^{eff}$ from 2 to $\infty$,
the tree-level dominant $\bar B^0\to p\bar p \pi^+\pi^-$ decay has
$\alpha^d_\pm \simeq V_{ub}V_{ud}^* a_2$ in Eq.~(\ref{amp1}) 
to be sensitive to the non-factorizable effects.
Since the non-factorizable effects are uncomputable, according to the data of 
${\cal B}(\bar B^0\to p\bar p \pi^+\pi^-)=(3.0\pm 0.3)\times 10^{-6}$~\cite{BtoBBMM_LHCb} 
in Eq.~(\ref{br_data}), we obtain 
${\cal B}(\bar B^0\to p\bar p \pi^+\pi^-)=(3.0\pm 0.9)\times 10^{-6}$,
where $a_2=0.26\pm 0.01$ 
 with the tiny value of $\delta a_2=0.01$ from the new data is compatible to
${\cal O}(0.2-0.3)$ from 
the two-body  $B$ and $\Lambda_b$ and  three-body baryonic $B$
decays~\cite{Neubert:2001sj,Hsiao:2015cda,Hsiao:2016amt}.
For the measured branching ratio of $\bar B^0\to p\bar p \pi^+ K^- +p\bar p \pi^- K^+$,
it is found that the contribution is mainly from 
the penguin-level dominant $\bar B^0\to p\bar p \pi^+ K^-$ mode.
Note that
$a_{3,5}$ from $\alpha^s_\pm \simeq \beta^s_\pm=-V_{tb}V_{ts}^*(a_3\pm a_5+a_9)$
are also sensitive to the non-factorizable effects. With $N_c^{eff}=3$, 
we obtain ${\cal B}(\bar B^0\to p\bar p \pi^+ K^-)=(6.6\pm 2.4)\times 10^{-6}$,
which suggests that the decay is free from the non-factorizable effects. 
In Table~\ref{br_BtoBBMM}  we have included the data to constrain the non-factorizable effects, 
which results in $\delta N_c^{eff}=0.06$.
We note that the two spectra in Fig.~\ref{spec} for 
$B^-\to \Lambda\bar p M_1M_2$ and $\bar B^0\to p\bar p M_1M_2$
present the threshold effects as the peaks around the threshold areas of 
$m_{\Lambda\bar p}\simeq m_\Lambda +m_{\bar p}$ and 
$m_{p\bar p}\simeq m_p+m_{\bar p}$, respectively, 
which are commonly observed 
in the three and four-body baryonic $B$ decays~\cite{BtoBBMM_LHCb,Wang:2007as}. 

Finally, we remark that 
we cannot explain the 
data of ${\cal B}(\bar B^0_s\to p\bar p K^\pm \pi^\mp, p\bar p K^+ K^-)
=(1.5\pm 0.7,4.6\pm 0.6)\times 10^{-6}$
measured by the LHCb~\cite{BtoBBMM_LHCb}
due to the lack of the information for the transition form factors of $\bar B^0_s\to (K^+\pi^-,K^+ K^-)$.
This calls for 
the theoretical and experimental studies of 
the three-body mesonic $\bar B^0_s$ decays that could proceed
with the $\bar B^0_s\to M_1 M_2$ transitions, 
such as the 
$\bar B^0_s\to D_s^{*-}\pi^+ K^0$,
$\bar B^0_s \to D^{*0}\pi^+ K^-(K^+K^-)$ and 
$\bar B^0_s\to \rho^- \pi^+ K^0$ decays with one of the mesons 
to be a vector one, 
in order to extract both $(h,w_-)$ in Eq.~(\ref{BtoMM}). On the other hand, 
the observed 
$\bar B^0_s\to D^0 K^+ \pi^-$ and $\bar B^0_s\to D^0 K^+ K^-$ decays~\cite{pdg}
are also important as they relate to $w_-$.

\section{Conclusions}
In sum, we have studied the charmless four-body baryonic $B\to {\bf B_1 \bar B_2}M_1 M_2$ decays,
where the primary decaying processes are regarded as 
the $B\to M_1M_2$ transitions along with the baryon-pair productions.
According to the new extractions of the $B\to M_1 M_2$ transition form factors from
the three-body $B\to D^{(*)}M_1 M_2$ and $B\to M_1M_2M_3$ decays,
we have shown that  
${\cal B}(B^-\to \Lambda\bar p \pi^+\pi^-)=(3.7^{+1.5}_{-1.0} )\times 10^{-6}$ and 
${\cal B}(\bar B^0\to p\bar p \pi^+\pi^-,p\bar p \pi^+ K^-)=
(3.0\pm 0.9,6.6\pm 2.4)\times 10^{-6}$, which agree with the data.
We have also predicted 
${\cal B}(B^-\to\Lambda\bar p K^+ K^-)=(3.0^{+1.3}_{-0.9})\times 10^{-6}$
to be accessible to 
 the LHCb and BELLE experiments.
The study of $B\to {\bf B_1 \bar B_2}M_1 M_2$ benefits the future test of T violation,
as the T-odd triple momentum product correlation of $\vec{p}_1\cdot (\vec{p}_2\times \vec{p}_3)$
can be directly constructed.

\section*{ACKNOWLEDGMENTS}
We would like to thank Dr. Eduardo Rodrigues for useful discussions.
This work was supported in part by National Center for Theoretical Sciences,
MoST (MoST-104-2112-M-007-003-MY3), and
National Science Foundation of China (11675030).

\end{document}